\documentclass[preprint,aps]{revtex4}               

\usepackage{graphicx}
\usepackage{dcolumn}
\usepackage{bm}
\usepackage{sidecap}

\begin{document}

\preprint{APS/123-QED}

\title{Late time tails of fields around Schwarzschild black hole surrounded by quintessence}

\author{Nijo Varghese}  
 \email{nijovarghese@cusat.ac.in}
\author{V C Kuriakose}%
 \email{vck@cusat.ac.in}
\affiliation{%
Department of Physics, Cochin University of Science and Technology,
Cochin - 682 022, Kerala, India.
}%


\begin{abstract}
The evolution of scalar, electromagnetic and gravitational fields around spherically symmetric black hole surrounded by quintessence are studied with 
special interest on the late-time behavior. In the ring down stage of evolution, we find in the evolution picture that the fields decay more 
slowly due to the presence of quintessence. As the quintessence parameter $\epsilon$ decreases, the decay of $\ell=0$ mode of scalar field 
gives up the power-law form of decay and relaxes to a constant residual field at asymptotically late times. The $\ell>0$ modes 
of scalar, electromagnetic and gravitational fields show a power-law decay for large values of $\epsilon$, but for smaller values of $\epsilon$ 
they give way to an exponential decay.
\end{abstract}

\maketitle

\section{Introduction}
\label{intro}
Black holes are among the simplest objects in the Universe as they can be fully 
described by merely three quantities viz., mass, charge and angular momentum. During the 
process of gravitational collapse of a charged rotating star leading to the formation of a black hole, the geometry 
outside the star relaxes to the stationary Kerr-Newman spacetime. This result, what is now known 
as \textquotedblleft {\it no-hair theorem}\textquotedblright\, means that any \textquotedblleft hair\textquotedblright
other than its mass, charge and angular momentum, will disappear after the collapsing body settles down to 
its stationary state. The mechanism responsible for the vanishing of these 
quantities remained unknown till Richard Price\cite{price} who, making a perturbative analysis of the 
collapse of a nearly spherical star, showed that for a field with spin, s, any radiative multipole 
($\ell\geq s$) gets radiated away completely, in the late stage of collapse. Further, he showed that at late 
times the field dies out with a power-law tail $t^{-(2\ell+p+1)}$, where $p=1$ if the multipoles were initially 
static and $p=2$ otherwise. Later, in\cite{gundlach} a detailed analytical analysis supported by numerical 
results of massless fields around Schwarzschild and Reissner-Nordstr\"om(RN) black hole showed that power-law tails are also present 
on null infinity where they decay as $u^{-(\ell+p)}$ and on future event horizon as $v^{-(2\ell+p+1)}$.

The studies of the evolution of charged scalar field around RN spacetime\cite{hod1} reveal that a charged 
hair sheds slower than a neutral hair. In contrast with the massless fields, massive fields with mass, $m$, have an
oscillatory inverse power-law behavior $t^{-(\ell+3/2)}sin(mt)$, at intermediate late times \cite{hod,NV2}. 
But it was shown that in the asymptotic late times another pattern of
oscillatory tail of the form, $t^{-(5/6)}sin(mt)$ dominates\cite{koyama,konoplya}. 
The late-time tails appearing in the propagation of fields in the higher dimensional black hole 
were studied in \cite{cordoso,abdalla}.

The evolution of fields propagating on spacetime with a non vanishing cosmological constant were 
addressed in\cite{brady3,chambers,brady1,brady2,molina} and they have demonstrated the existence of exponentially decaying tails 
at late times contrasting the power-law tails in asymptotically flat situation. 

Earlier studies provide us the following picture of evolution of perturbations in the black hole spacetimes.
It involves three stages\cite{nollert}, the first one is an initial response, determined by the particular form of
the original wave field followed by a region dominated by damped oscillation of the
field called quasinormal modes(QNMs), which depend entirely on the background black hole
spacetimes. The QNMs play an important role in almost all astrophysical processes
involving black holes and the spectra of QNMs are expected to be detected by the future gravitational wave detectors. At late-times 
the QNMs are suppressed by the so called \textit{tail} form of decay. This comprises the last stage of evolution and is an interesting 
topic of study, since it reveals the actual physical mechanism by which a perturbed black hole sheds its hairs.

In the past two decades there have been growing observational evidences\cite{perl,riess} which indicate clearly that our universe is 
expanding in an accelerated pace, indicating the presence of some mysterious form of repulsive energy called dark energy. 
In order to explain the nature of dark energy, several models have been proposed. The simplest option being Einstein's cosmological
constant, $\Lambda$, which has a constant equation of state with state parameter, $\epsilon=-1$, but it needs extreme fine tuning to account
for the observations\cite{weinberg}. Models were proposed, replacing $\Lambda$ with a dynamical,
time-dependent and spatially inhomogeneous scalar field now called quintessence,
which can have an equation of state, $-1\leq\epsilon\leq-1/3$\cite{ratra,caldwell}.

Spherically symmetric black hole solutions\cite{kiselev} were obtained in quintessence model of dark energy and many 
authors\cite{massless,qqnm2,qqnm3,qqnm4,qqnm5,qqnm6,qqnm7,qqnm8,qqnm9} considered the evolution of various fields around 
black holes surrounded by quintessence. All these studies are based on the calculation of the 
QNMs using the third order WKB approximation method while the late-time behavior remained unexplored till now. 
In fact the late-time decay is determined by the asymptotic curvature of the spactime, so it is interesting to see how the 
fields evolve in a spacetime in which the asymptotic structure is determined by the quintessence field and our study fills this gap. 

The rest of the paper is organized as follows. In Sect.\ref{sec2} we introduce the master wave equation for
massless scalar, electromagnetic and gravitational field perturbations around black hole surrounded by quintessence. The numerical method used to
study the time evolution is explained in Sect.\ref{sec3} and the results are presented. The conclusions are given in Sect.\ref{sec4}.

\section{Fields around black hole surrounded by Quintessence}
\label{sec2}
The exact solution of Einstein's equation for a static spherically symmetric black hole surrounded by the quintessential matter,
under the condition of additivity and linearity in energy momentum tensor, was found in\cite{kiselev},

\begin{equation}
\ ds^{2}=-f(r)dt^{2}+f(r)^{-1}dr^{2}+r^{2}(d\theta ^{2}+\sin^{2}\theta d\phi ^{2}), 
\label{metric}
\end{equation}
where $f(r)=\left(1-\frac{2M}{r}-\frac{c}{r^{3\epsilon+1}}\right)$, $M$ is the mass of the black hole, $\epsilon$ is the 
quintessential state parameter and $c$ is the normalization factor which depends on the density
of quintessence as, $\rho_{q}=\frac{-c}{2}\frac{3\epsilon }{r^{3(1-\epsilon )}}$.
It is difficult to analyze the above metric for arbitrary values of the parameter $\epsilon$.
For our study we take three special cases of the quintessence parameter, $\epsilon=-1/3,-2/3$ and $-1$ along with the Schwarzschild(Schw)
case($c=0$), so that the calculations become viable.

{\bf Case1 $\epsilon=-1/3$ :}
the black hole event horizon is located at $r_{e1}=2M/(1-c)$, corresponding to zero of the function, $f(r)$ and 
the tortoise coordinate can be defined as,
\begin{equation}
r_{*}=r+r_{e1}ln(r-r_{e1}).
\label{tort1}
\end{equation}

{\bf Case2 $\epsilon=-2/3$ :} here, in addition to the black hole event horizon at $r=r_{e2}$ the 
spacetime possesses a cosmological horizon at $r=r_{c2}$, with $r_{e2}<r_{c2}$. In terms of these horizons we 
can write $f(r)$ as, 
\begin{equation}
f(r)=\frac{c}{r}(r-r_{e2})(r_{c2}-r). 
\end{equation}

The surface gravity associated with the horizons at $r=r_{i}$, is defined by 
\begin{equation}
\kappa_{i}=\frac{1}{2}|df/dr|_{r=r_{i}},
\end{equation}
 
and we get,
\begin{equation}
\kappa_{e2}=\frac{c(r_{c2}-r_{e2})}{2r_{e2}}, \quad\quad  \kappa_{c2}=\frac{c(r_{e2}-r_{c2})}{2r_{c2}}. 
\end{equation}

These quantities allow us to write,

\begin{equation}
\frac{1}{f}=\frac{1}{2\kappa_{e2}(r-r_{e2})}+\frac{1}{2\kappa_{c2}(r-r_{c2})}.
\end{equation}

Now we can get an expression for the tortoise coordinate, $r_{*}=\int f^{-1}dr$ as,
\begin{equation}
 r_{*}=\frac{1}{2\kappa_{e2}}ln\left|\frac{r}{r_{e2}}-1\right|+\frac{1}{2\kappa_{c2}}ln\left|1-\frac{r}{r_{e2}}\right|.
\label{tort2}
\end{equation}

{\bf Case3 $\epsilon=-1$:}  this is the extreme case of quintessence, the Schwarzschild-de Sitter(SdS) spacetime. 
The roots of the polynomial equation, $f(r)=0$ corresponding to the event horizon at $r=r_{e3}$, the cosmological 
horizon at $r=r_{c3}$ and a negative root at $r=r_{0}=-(r_{e3}+r_{c3})$, with $r_{0}<r_{e3}<r_{c3}$.  
We can define the tortoise relation in terms of these horizons and the corresponding surface gravity as\cite{brady2},

\begin{equation}
r_{*} = \frac{1}{2\kappa_{e3}}ln\left|\frac{r}{r_{e3}}-1\right|-\frac{1}{2\kappa_{c3}}ln\left|1-\frac{r}{r_{c3}}\right|+\frac{1}{2\kappa_{0}}ln\left|\frac{r}{r_{0}}-1\right|.
\label{tort3} 
\end{equation}

The evolution of massless scalar field $\Phi$, electromagnetic(EM) field $A_{\mu}$ and gravitational perturbations(GR) $h_{\mu\nu}$ 
in the spacetime $g_{\mu\nu}$, specified by Eq.(\ref{metric}) are governed by the Klein-Gordon\cite{massless}, 
Maxwell\cite{ruffini} and Regge-Wheeler\cite{regwhe,qqnm3} equations, respectively,

\begin{equation}
\frac{1}{\sqrt{-g}}\partial_{\mu}(\sqrt{-g}g^{\mu\nu}\partial_{\nu})\Phi-m^{2}\Phi=0,
\label{KGeqn}
\end{equation}
\begin{equation}
F^{\mu\nu}_{;\nu}=0, \quad\quad with\quad F_{\mu\nu}=A_{\nu,\mu}-A_{\mu,\nu}
\label{Max}
\end{equation}
\begin{equation}
 R_{\mu\nu}(g+h)=0,   \label{GR}
\end{equation}

where $R_{\mu\nu}(g+h)$ is the Ricci tensor calculated by considering small perturbations, $h_{\mu\nu}$, in the 
background spacetime $g_{\mu\nu}$. For gravitational field, there are two modes of perturbations, axial and polar. 
Here we are considering only the axial perturbation for our study.
The two were shown to have a similar QNM spectra and late time behavior in de Sitter spacetime\cite{molina}. 
The radial part of all the above perturbation equations can be decoupled from their angular parts and reduced to the form,

\begin{equation}
\left(-\frac{\partial^{2}}{\partial
t^{2}}+\frac{\partial^{2}}{\partial
r^{2}_{*}}\right)\psi_{\ell}(t,r)=-V_{\ell}(r)\psi_{\ell}(t,r),
\label{waveqn}
\end{equation}

where the tortoise coordinate is defined by, $dr_{*}=\frac{1}{f}dr$ and the effective potentials are given by,

\begin{eqnarray}
V_{SC} &=& f(r)\left (\frac{\ell(\ell+1)}{r^{2}}+\frac{2M}{r^{3}}+\frac{c(3\epsilon+1)}{r^{3\epsilon+3}}\right), \\
V_{EM} &=& f(r)\left (\frac{\ell(\ell+1)}{r^{2}}\right), \\
V_{GR} &=& f(r)\left (\frac{\ell(\ell+1)}{r^{2}}-\frac{6M}{r^{3}}-\frac{c(3\epsilon+3)}{r^{3\epsilon+3}}\right).
\end{eqnarray}

\section{Numerical integration and results}
\label{sec3}
The complex nature of the potentials makes it difficult to obtain the
exact solutions of Eq.(\ref{waveqn}) and we have to tackle the
problem by numerical methods. A simple and efficient method to study the evolution of field were developed in\cite{gundlach},
 after recasting the wave equation, Eq.(\ref{waveqn}), in the null coordinates, $u=t-r_{*}$ and $v=t+r_{*}$ as,
\begin{equation}
-4\frac{\partial^{2}}{\partial u \partial v}\psi(u,v)=V(u,v)\psi(u,v). \label{waveqn2}
\end{equation}

Using the following discretization,

\begin{equation}
\psi_{N}=\psi_{W}+\psi_{E}-\psi_{S}-\frac{h^{2}}{8}V(S)(\psi_{W}+\psi_{E})+O(h^{4}),
\end{equation}

we perform the numerical integration on an uniformly spaced grid with points, $N(u+h,v+h)$, $W(u+h,v)$, $E(u,v+h)$ and $S(u,v)$ forming a null
rectangle with an overall grid scale factor of $h$. Since the late-time behavior of the wave function is found to be insensitive
to the initial data, we set $\psi(u,v=0)=0$ and a Gaussian profile, $\psi(u=0,v)=A exp\left [-\frac{(v-v_{0})^2}{2\sigma^2} \right ]$.
In all our calculations we set the initial Gaussian with width, $\sigma=3$ centered at $v_{0}=10$. Due to the linearity of equation
one has the freedom to choose the amplitude of the initial wave $A=1$. To proceed the integration on the above numerical scheme one
has to find the value of the potential at $r(r_{*})=r((v-u)/2)$ at each step. Employing the Newton-Raphson method one can invert 
Eqs.(\ref{tort1}), (\ref{tort2}) and (\ref{tort3}) and get $r(r_{*})$ for the characteristic integration. 

\begin{figure}[h]
  \includegraphics[width=0.45\textwidth]{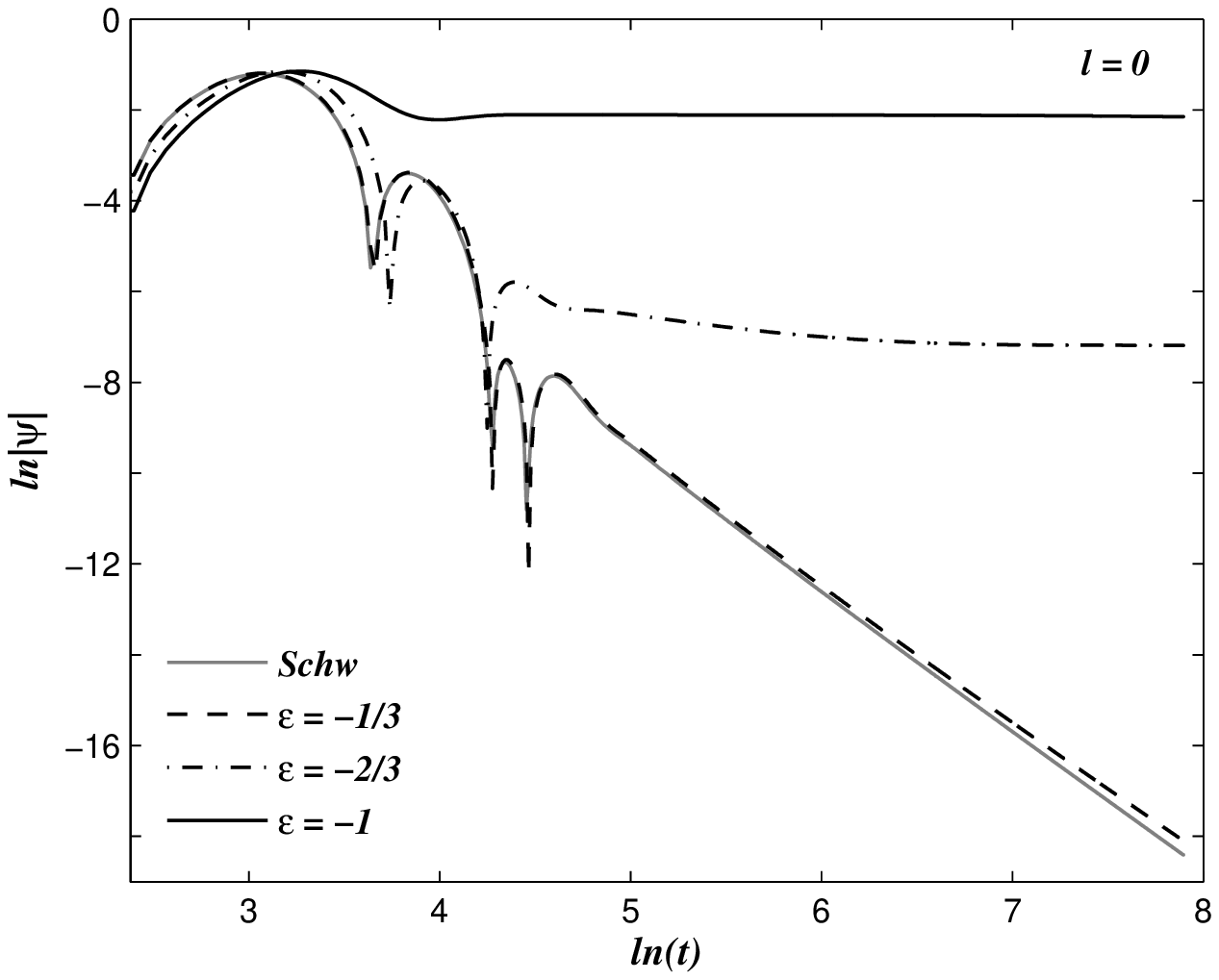}
  \includegraphics[width=0.45\textwidth]{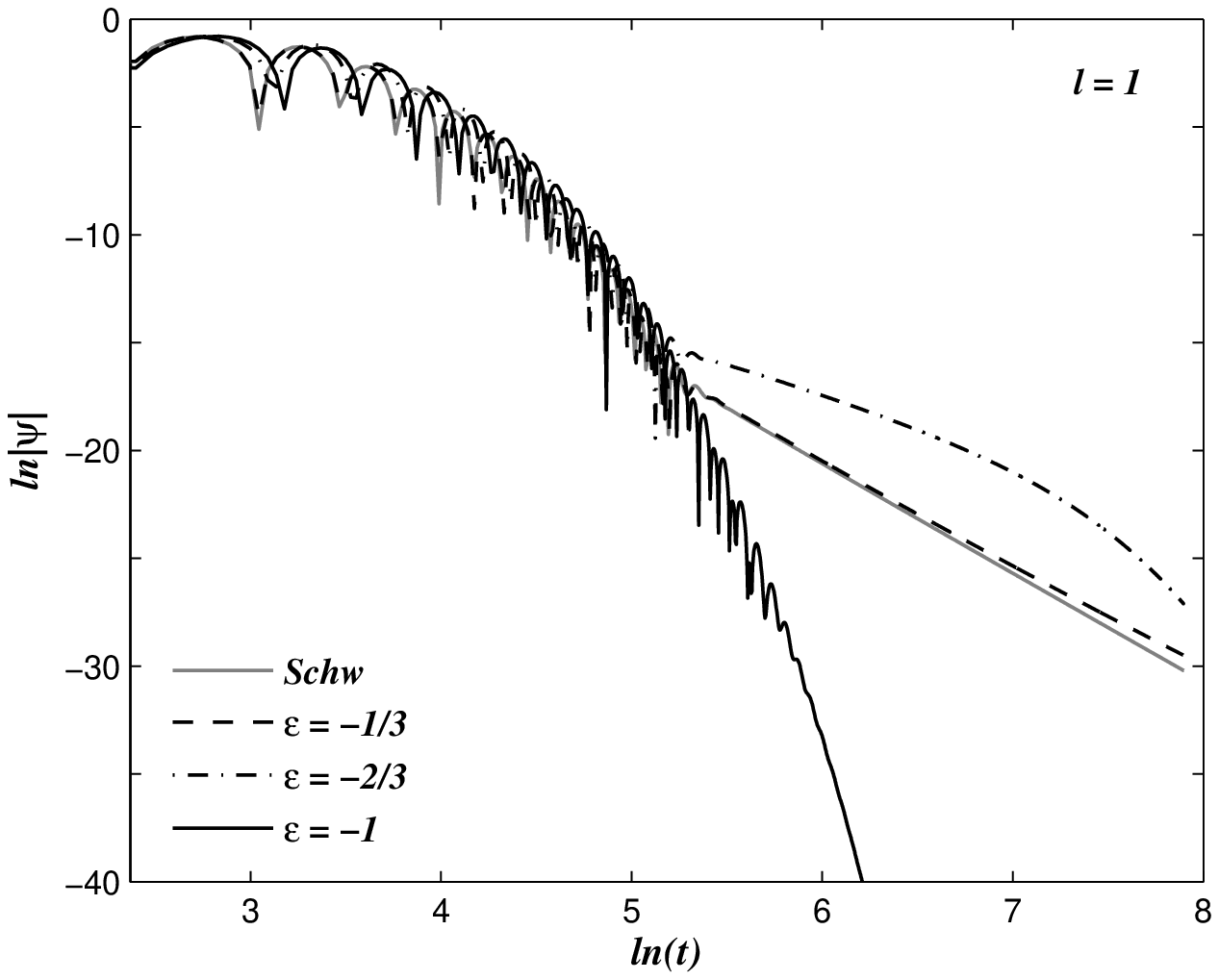}
\caption{Log-log graph of the evolution of scalar field in a quintessence filled black hole spacetime with quintessence 
parameters $\epsilon=-1/3,-2/3,-1$ and $c=10^{-2}/2$, in comparison with that in the pure Schwarzschild spacetime. 
The case of $\ell=0$ mode is shown in the left plot and $\ell=1$ on the right.}
\label{sc1}
\end{figure}

Figure \ref{sc1} shows typical evolution profile of scalar field in a 
quintessence surrounded black hole, in comparison with that in the pure Schwarzschild spacetime. We observe that the evolution shows 
deviations from the Schwarzschild case, after initial transient phase. The damped single frequency oscillation(QNM) phase and the 
late-time tail of decay in the final phase shows the characteristics of the quintessence. We analyze each phase in detail.

\subsection{Quasinormal modes}

\begin{figure}[h]
\includegraphics[width=0.45\columnwidth]{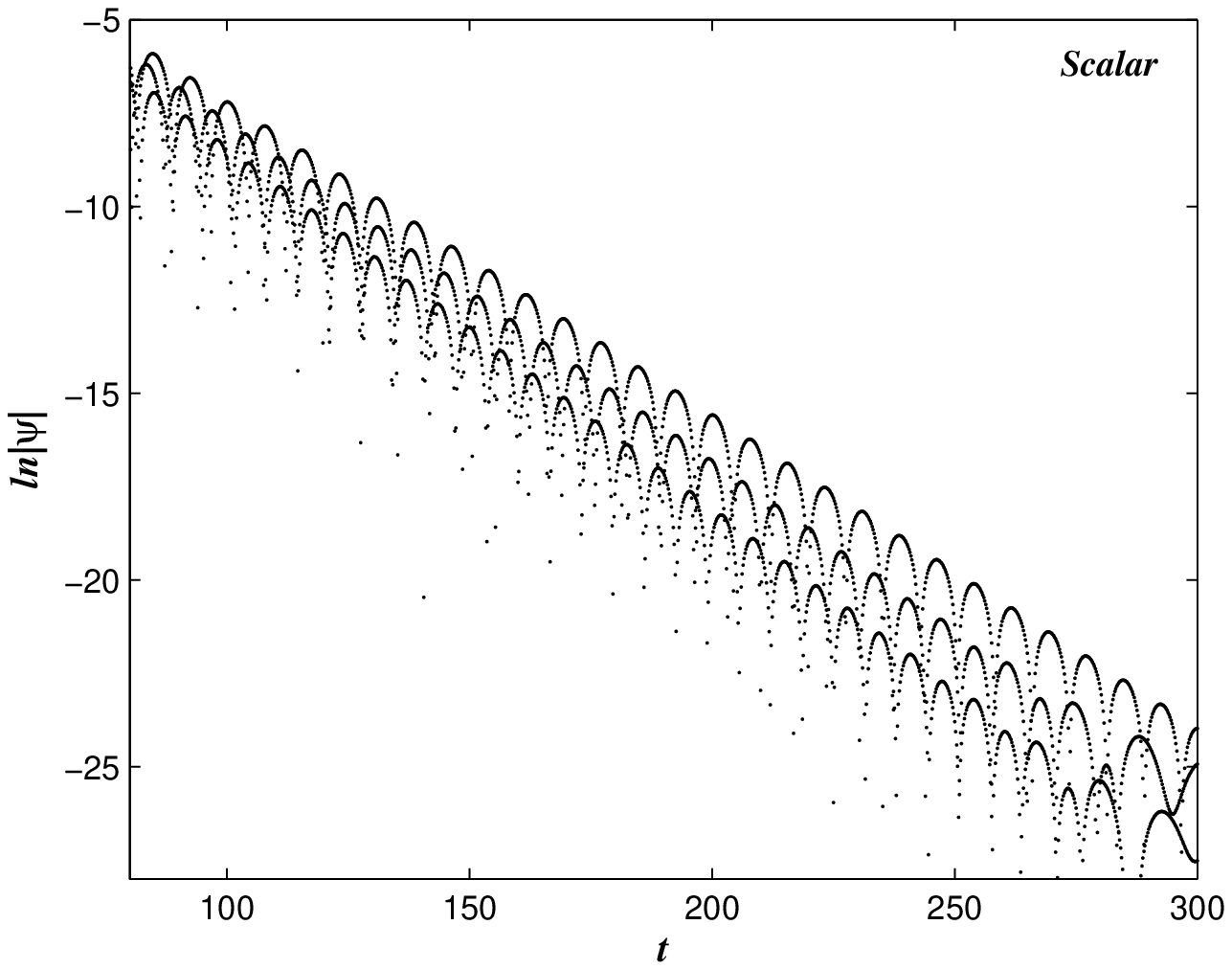}
\includegraphics[width=0.45\columnwidth]{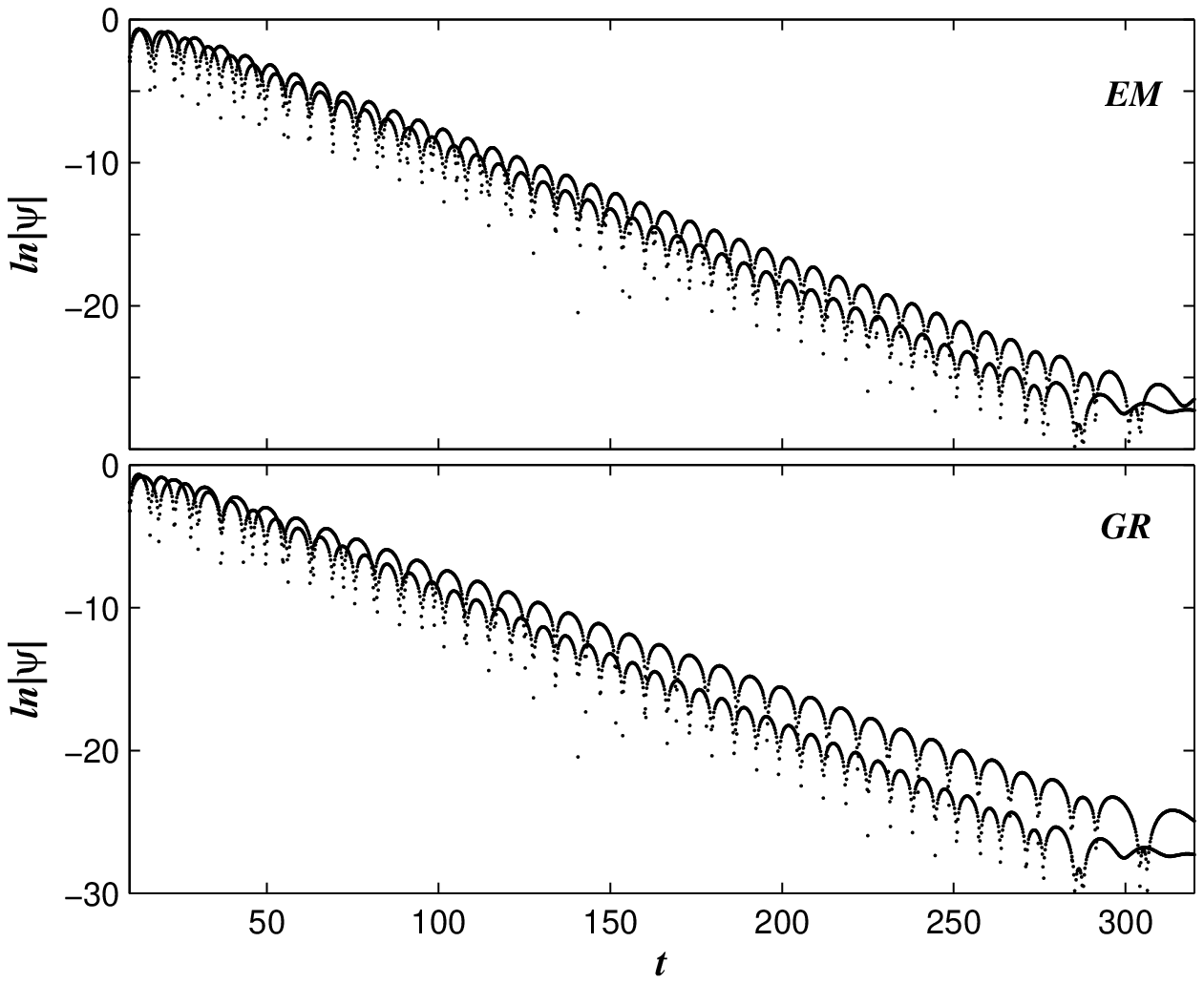}
\caption{Semi-log graph of the QNM phase of evolution for the $\ell=2$ modes of various fields. The plot on the left is for scalar field where the curves 
from bottom to top is for the Schwarzschild, $\epsilon = -2/3$ and -1  cases, respectively.  
On the right EM and GR fields are shown where bottom curve in each sub plot represents the Schwarzschild case and the other ones are for $\epsilon = -2/3$.
We take $c=10^{2}/2$ for these plots. All the three fields damp more slowly in the presence of quintessence.} \label{QNM}
\end{figure}

QNMs of various field perturbations around black hole surrounded by quintessence were evaluated in 
\cite{massless,qqnm2,qqnm3,qqnm4,qqnm5,qqnm6,qqnm7,qqnm8,qqnm9} using third order WKB method. We are not going for a quantitative 
study of QNMs but the typical nature of QNM phase can be seen in Figure \ref{QNM},
where we have plotted the evolution profiles of the different fields in a quintessence surrounded black hole for $\ell = 2$, in 
comparison with the corresponding mode in the pure Schwarzschild black hole. On the left pannel the decay of scalar field is shown for 
$\epsilon=-2/3$ and $-1$ along with the Schwarzschild case.

It can be seen that the decay of the fields in the QNM phase is slower in the pressence of quintessence than in the pure Schwarzschild cases. Also 
the damping is lower for smaller values of $\epsilon$.
This observation is in contrast with the results presented in \cite{massless}, where thay obtained that a massless scalar field damps more rapidly 
due to quintessence. Electromagnetic and gravitational field perturbations also show similar behavior as that of scalar field, in the QNM phase. 
So we have shown the $\epsilon=-2/3$ case only in the plots. These results are in agrement with the observations made in 
\cite{qqnm2,qqnm3,qqnm4}, using the WKB method. 

\subsection{Late-time tails}
The QNM phase is followed by the regime of late-time tails of field decay. We find that the nature of late-time tails are sensitive to the 
parameters of quintessence, the angular momentum index and also the type of field perturbations under consideration. 
Below we discuss the different cases in detail.

\subsubsection{$\ell=0$ mode}

Scalar field perturbations can have the $\ell=0$ mode, while EM and GR fields can only have modes with $\ell\geq1$ and $\ell\geq2$ respectively. 
From Figure \ref{sc1}, which is a log-log plot, it can be seen that for the $\ell=0$ mode, scalar field with $\epsilon=-1/3$ case of quintessence 
has the form of a power-law tail, with slightly slower decay rate than the corresponding Schwarzschild tail.

\begin{SCfigure}[][h]
\centering
\includegraphics[width=0.6\columnwidth]{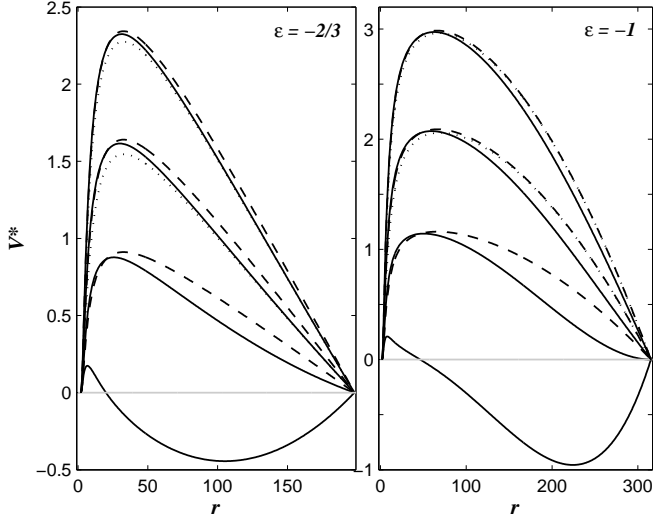}%
\caption{Plot of effective potentials for $\epsilon = -2/3$ with $c=10^{-2}/2$ and $\epsilon = -1$ with $c=10^{-5}$. Curves from bottom to top is 
for $\ell=0,1,2$ and 3 modes. The scalar (solid curves), EM (dashed) and GR(dotted) fields are drawn.
At large distances, GR potential is merged with the scalar case for $\epsilon = -2/3$ and with EM case for $\epsilon = -1$. The potential is scaled as $V^{*}=V(r_{e}-r)^{2}(\ell+1/2)$, to enhance the nature at large r.} 
\label{potplot}
\end{SCfigure}

For $\epsilon=-1/3$, the effective potential of scalar field is positive definite for $r_{*}\in[-\infty,+\infty]$ and have a potential 
barrier near the event horizon but vanish asymptotically as  $r_{*}\rightarrow\pm\infty$. But for smaller values of the quintessence parameter, 
$\epsilon=-2/3$ and -1, after a barrier nature near the event horizon, the 
potentials with $\ell=0$ mode of scalar field, vanish at some $r_{*}=r_{*}^{0}$ and there after form a negative 
well in the range $r_{*}^{0}<r_{*}<+\infty$. Figure \ref{potplot} shows potentials experienced by the scalar, EM and GR fields.
To clearly see the behavior of the potential between the horizons, we choose the value of the parameter $c=10^{-5}$ for $\epsilon=-1$ case 
and a larger value of $c=10^{-2}/2$ for $\epsilon=-2/3$ case, so that we can bring down the seperation between the horizons.

\begin{SCfigure}
\centering
\includegraphics[width=0.65\columnwidth]{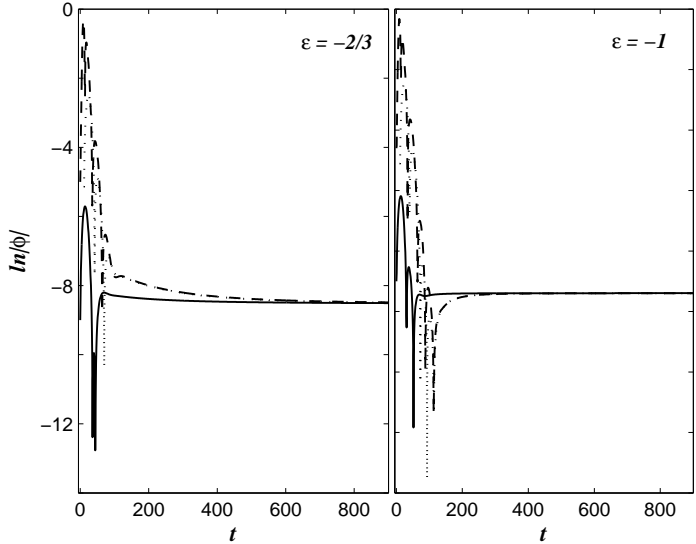}%
\caption{The decay of $\ell=0$ mode of scalar field with quintessence parameters $\epsilon = -2/3$ with $c = 10^{-2}/2$, 
and $\epsilon = -1$ with $10^{-5}$. The fields are evaluated on the constant surface $r_{*}=10$(dotted curves), black hole event horizon(dashed curves) 
and cosmological horizon(solid curves).}
\label{3surf}
\end{SCfigure}

This peculiar shape of potentials for $\epsilon=-2/3$ and -1 reflects in the late-time tails, that 
the fields settle to a residual constant value, rather than decaying to zero. We have evaluated the field, $\phi=\psi/r$ on the surface 
of constant r(at $r_{*}=10$), the cosmological horizon(approximated by the null surface $v=v_{max}$) and the black hole event horizon(approximated by the null 
surface $u=u_{max}$). In the asymptotic late times field settles to the same constant value, $\phi_{0}$ on all these surfaces, as shown in 
Figure \ref{3surf}. It was first observed in\cite{brady1}, for scalar field in the SdS spacetime and they have shown that at late times,

\begin{equation}
\phi\mid_{\ell=0}\simeq\phi_{0}+\phi_{1}(r)e^{-2k_{c3}t}.
\label{contfield}
\end{equation}

\begin{figure}[h]
\includegraphics[width=0.45\columnwidth]{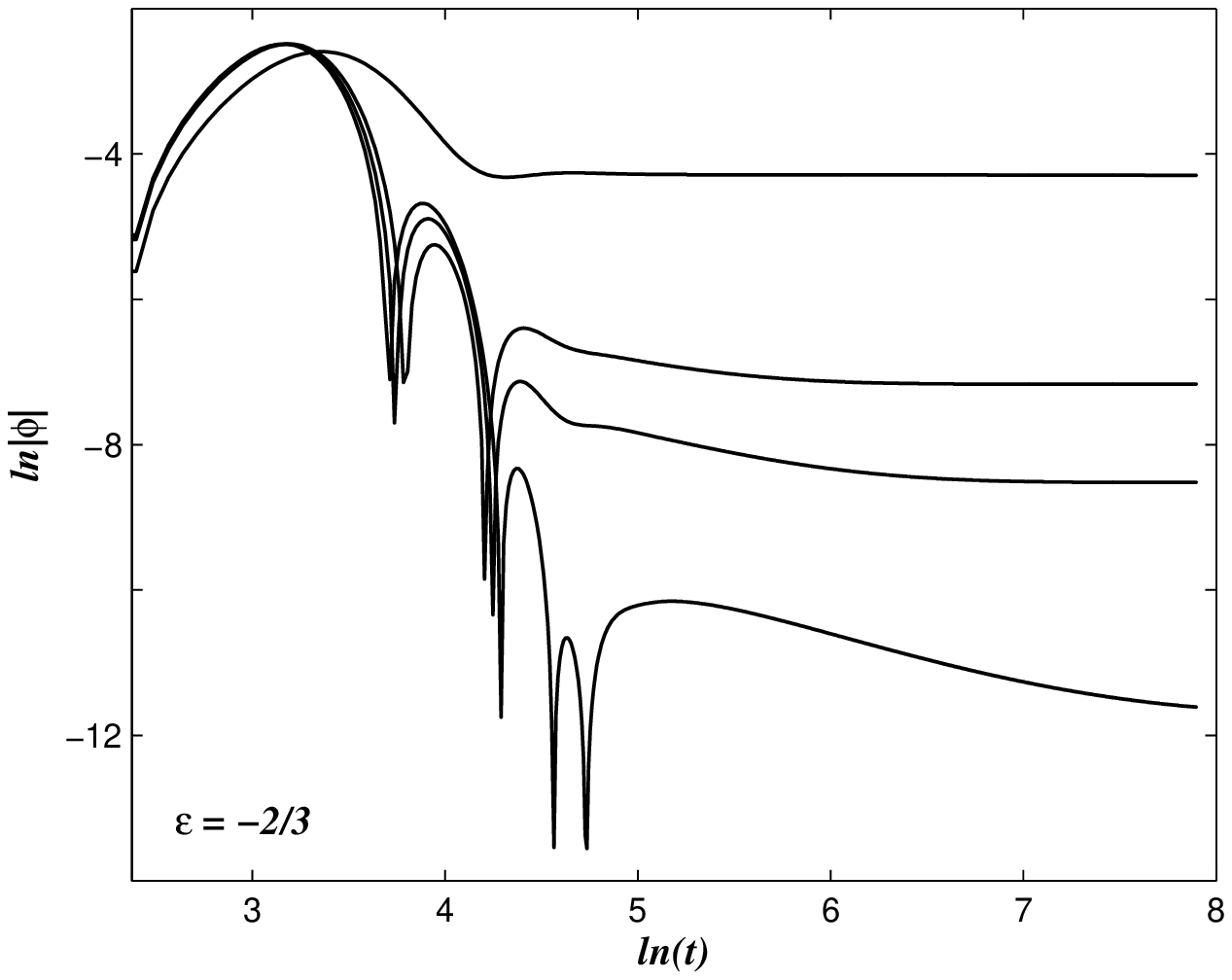}
\includegraphics[width=0.45\columnwidth]{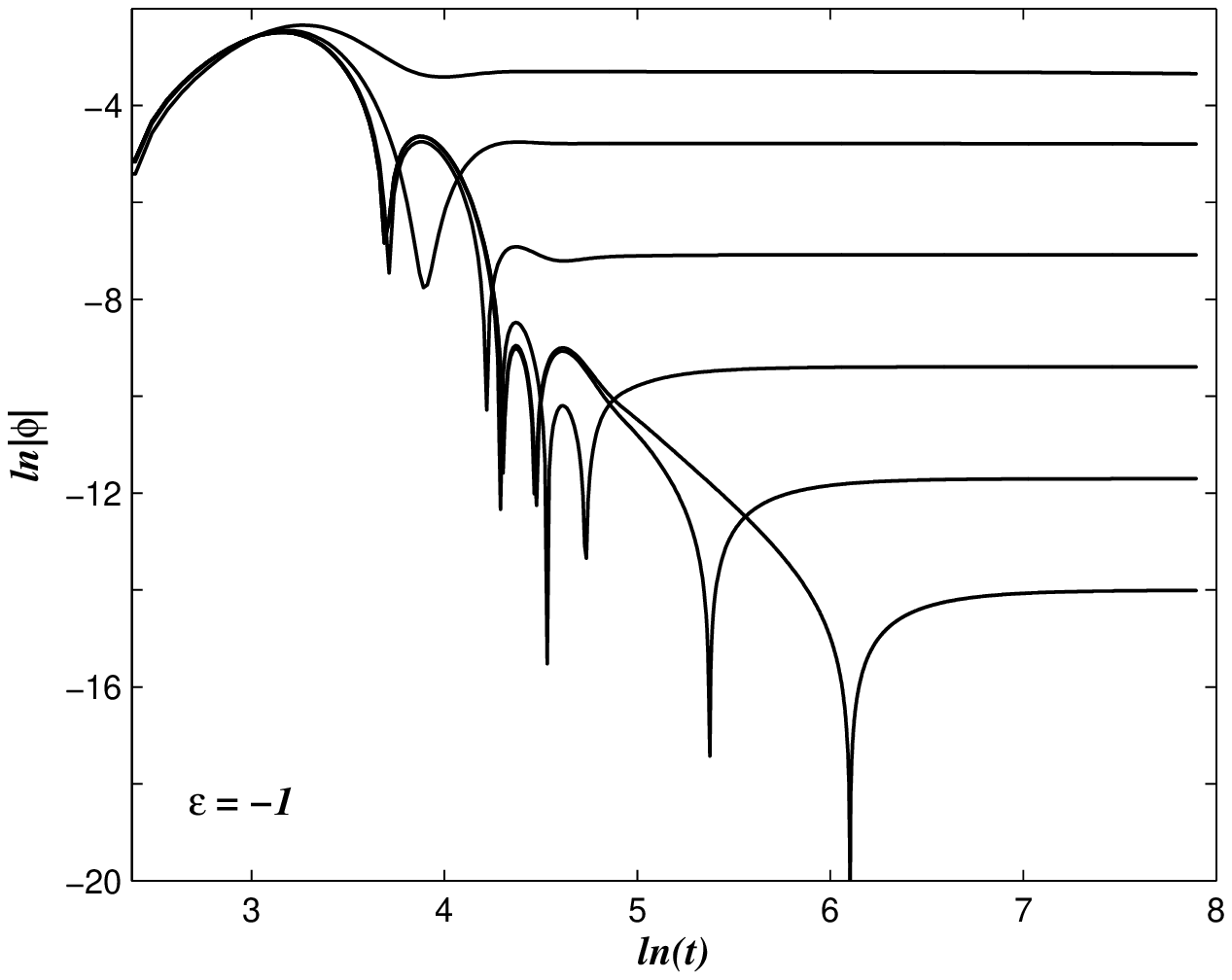}
\caption{The decay of $\ell=0$ mode of scalar field. On the left pannel, the $\epsilon = -2/3$ case is shown where the curves from top to bottom represents the 
plots for $c = 10^{-1}/2,10^{-2},10^{-2}/2$ and $10^{-3}$, respectively.  The $\epsilon = -1$ case is shown on the right, where the curves from top 
to bottom represents the plots for $c = 10^{-2}/2,10^{-3},10^{-4},10^{-5},10^{-6}$ and $10^{-7}$, respectively.}
\label{ScL0}
\end{figure}

\begin{figure}
\centering
\includegraphics[width=0.5\columnwidth]{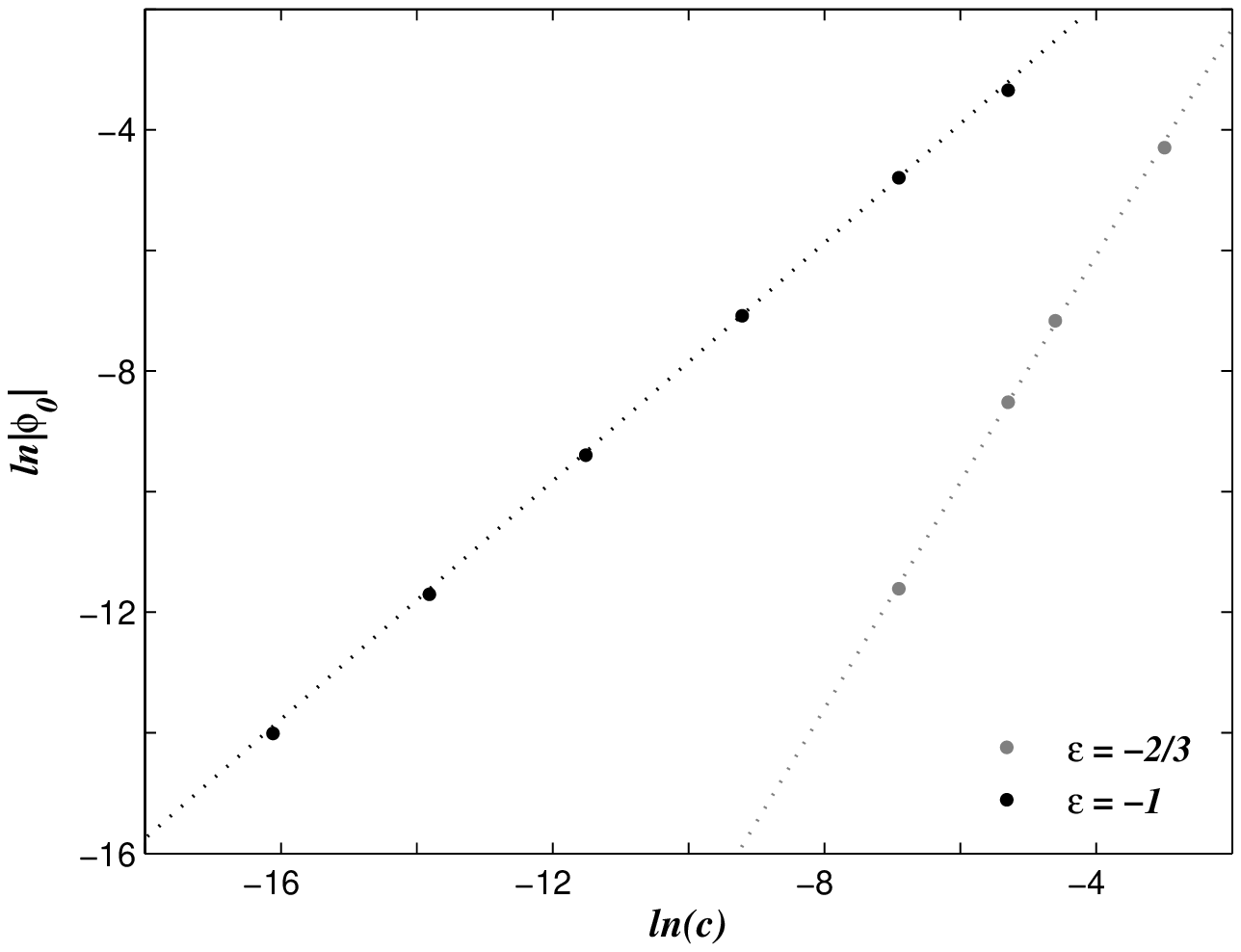}
\includegraphics[width=0.48\columnwidth]{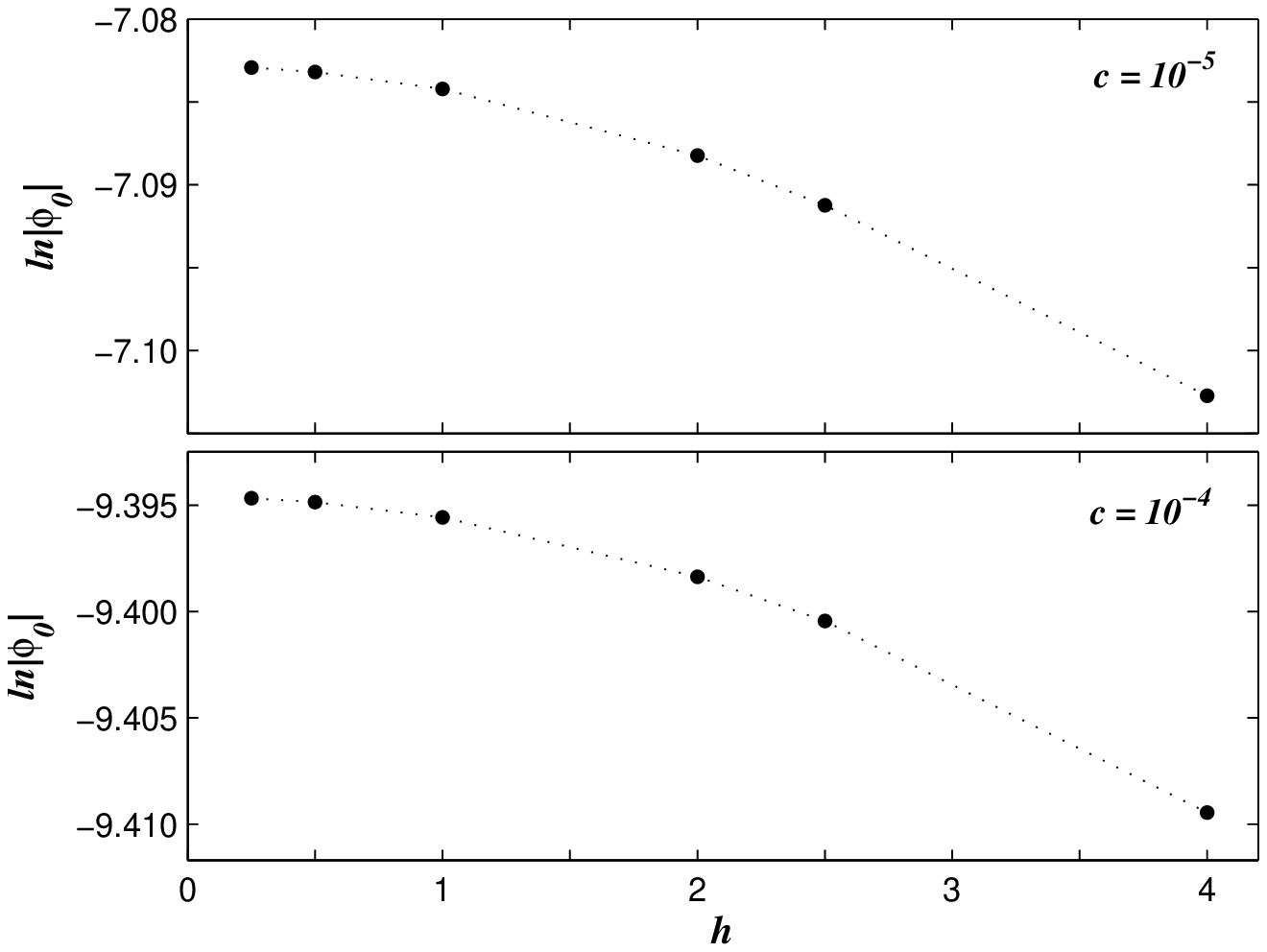}
\caption{On the left pannel, the asymptotic value of scalar field, $\phi_{0}$ versus c is plotted in logarithmic scale. A linear fit(dotted line) gives the 
slopes 1.873, for $\epsilon = -2/3$ and 0.995, for $\epsilon = -1$. On the right $ln|\phi_{0}|$ for different numerical grid resolution is plotted 
for the $\epsilon = -1$ case.}
\label{L0vsc}
\end{figure}

Figure \ref{ScL0} demonstrates the $\ell=0$ mode of scalar field for different values of c. The smaller the value of c, the later the 
field descends to the constant value, $\phi_{0}$. The dependence $\phi_{0}$, 
on the parameter $c$ is shown in Figure \ref{L0vsc}. We observe that the asymptotic value of field scales as 
$\psi_{0}\sim c^{ 1.873}$, when $\epsilon=-2/3$ and $\psi_{0}\sim c^{0.995}$, when $\epsilon=-1$ as found in\cite{brady1}. The possible 
numerical errors in these results can be visualized from the right plot in Figure \ref{L0vsc}. The error increases quickly as the 
grid scale, h increases.

\subsubsection{$\ell>0$ modes}
\begin{SCfigure}[][h]
\centering
\includegraphics[width=0.6\columnwidth]{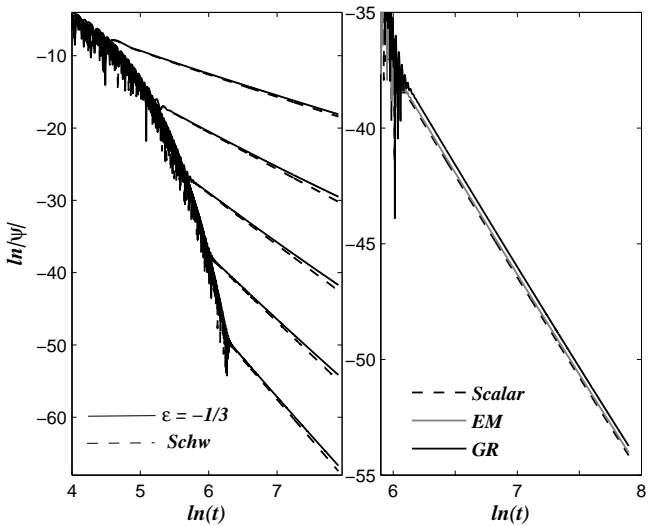}%
\caption{Log-log plot of the evolution of fields for $\epsilon=-1/3$, with $c=10^{-2}/2$. On the left decay of scalar field is shown 
in comparison with the Schwarzschild case. Curves from top to bottom is for $\ell=0,1,2,3$ and 4. The decay has a lower power-law 
constant, $t^{-(2\ell+2.7)}$ for quintessence case. The $\ell=3$ mode of different fields is shown on the right. All the field has an 
identical tail of decay for $\epsilon=-1/3$.} 
\label{E13}
\end{SCfigure}

Now we report the results obtained for the decay of $\ell>0$ modes of field perturbations. Figure \ref{E13} demonstrates the evolution of scalar 
and electromagnetic and gravitational fields around a black hole surrounded by quintessence with $\epsilon = -1/3$, along with pure Schwarzschild case. The fields, 
$\psi$ are evaluated on the surface $r_{*}=10$. A straight line in such a plot shows a power-law tail. The late-time tails follow the power-law 
decay for $\epsilon = -1/3$, but with a smaller decay rate, than the Schwarzschild case of, $\psi\sim t^{-(2\ell+3)}$. For the quintessence case 
with $c=10^{-2}/2$, we get $\psi\sim t^{-(2\ell+2.7)}$.

\begin{SCfigure}[][h]
\centering
\includegraphics[width=0.6\columnwidth]{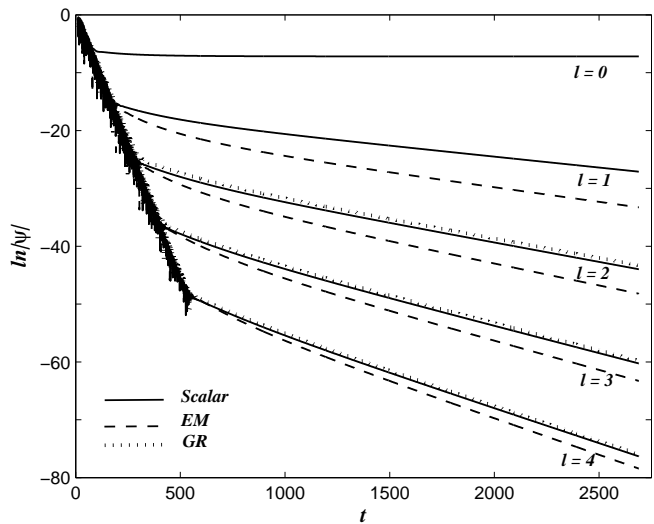}%
\caption{Semi-log plot of the decay of scalar and electromagnetic and gravitational perturbations for $\epsilon = -2/3$ with $c=10^{-2}/2$. GR has an 
identical decay pattern as that of scalar field.} 
\label{E23} 
\end{SCfigure}

\begin{SCfigure}[][h]
\centering
\includegraphics[width=0.6\columnwidth]{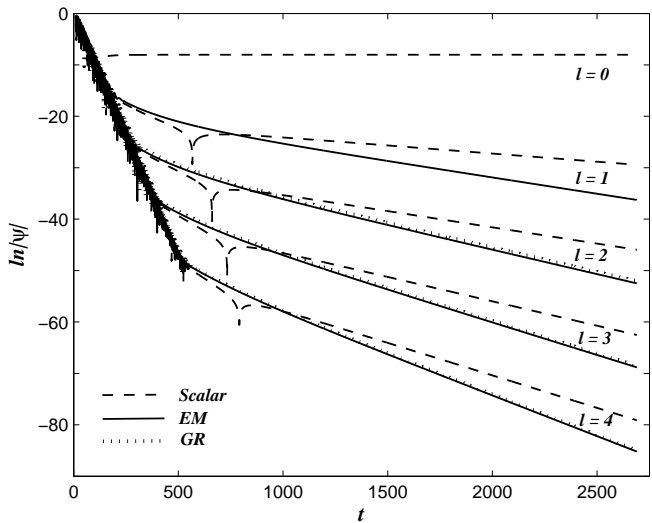}
\caption{Semi-log plot of the decay of scalar and electromagnetic perturbations for $\epsilon = -1$ with $c=10^{-5}$. The gravitational pertubations are 
merged with the EM profiles} 
\label{DeS}
\end{SCfigure}

As the value of the quintessence parameter decreases to $\epsilon=-2/3$ and -1, though in the intermediate time the fields 
follow the power-law decay, in the asymptotic late-time it deviates from the power-law form. 
Figure \ref{E23} is a semi-log plot of the decay of scalar, electromagnetic and gravitational perturbations for the $\epsilon=-2/3$ case 
with $c=10^{-2}/2$. A straight line in such a plot corresponds to an exponential decay. We can see from the plot that for $\ell>0$, the 
intermediate time power-law form of decay is replaced by an exponential decay of field, in the asymptotic late-time. In particular, in 
the asymptotic late times, the decay can be fitted in the forms,

\begin{eqnarray}
\psi & \sim & e^{-p k_{c2}\ell t}, \qquad\qquad \ \ for \ \ \ Scalar \ \ and \ \ GR, \nonumber \\
\psi & \sim & e^{-k_{c2}(\ell+1)t}, \qquad \ \ \ \ for \ \ \  EM, 
\label{E23a}
\end{eqnarray}

where p is some constant, for the plots shown in Figure \ref{E23} with $c=10^{-2}/2$ we get $p= 1.089$.
The evolution of field for the $\epsilon=-1$ case is plotted in Figure \ref{DeS}, in semi-log scale, where we have chosen $c=10^{-5}$, in order to 
see the decay for large $\ell$. All the $\ell>0$ modes at late-time relax as an exponential tail, according to the form, 

\begin{eqnarray}
\psi & \sim & e^{-k_{c3}\ell t}, \qquad \ \ \ \ \ for \ \ \ Scalar, \nonumber \\
\psi & \sim & e^{-k_{c3}(\ell+1)t}, \qquad for \ \ \  EM \ \ and \ \ GR. 
\label{D}
\end{eqnarray}

The late-time tails in black hole spacetimes are understood as a result of particular form the effective potential 
far away from the black hole. The tails are generated by the scattering of the waves off by the effective potential at great distances. 
Performing a systematic study of various potentials Ching {\it et al.}\cite{ching} and later by Hod\cite{hod2} had given the 
following heuristic picture. The late-time tails reported by an observer at $r_{*}$ are caused by the waves that are scattered by 
the potential, $V(r'_{*})$ at the point $r'_{*}\gg r_{*}$ and at late times, $\psi \propto V(r'_{*})\simeq V(t/2)$. Their studies show 
that the inverse power-law decay is a property of asymptotically flat spacetimes, and that different behaviors should be expected 
in black-hole spacetimes that have different asymptotic properties. The results that we obtained can be explained in this view. 
For $\epsilon=-1/3$, all the fields have inverse power nature of the effective 
potentials as that in asymptotically flat spacetime and lead to inverse power-law decay of fields at late times. 
When $\epsilon=-2/3$ and -1 the potentials exponentially drop off in the asymptotic region, $V(r_{*})\sim e^{-k_{c}r_{*}}$ and one can 
expect an exponential tail of the forms Eq.(\ref{E23a}) and (\ref{D}) at late times.

\section{Concluding Remarks}
\label{sec4}
Assuming a quintessence model of dark energy as the cause for the observed accelerated expansion of the universe, the paper studies the 
evolution of various field perturbations, especially the late-time decay of wave fields around a black hole spacetime. We find in the 
evolution picture that in the QNM stage, all the fields decay more slowly due to the presence of quintessence. At late times, QNMs are suppressed by the 
tail form of field decay. As the value of the quintessential parameter $\epsilon$, decreases, the late-time decay of $\ell=0$ mode of scalar field 
gives up the power-law form of decay, relaxing to a constant residual field. This asymptotic value of the field is determined by the 
value of the parameters, $\epsilon$ and c.

For the behavior of $\ell=0$ mode when $\epsilon=-2/3$ and -1, there is no analogue case in black holes with asymptotically flat spacetime. 
This can be understood as a consequence of cosmological no-hair theorem\cite{chambers}. Comparing the situation in SdS spacetime with the 
scattering inside the charged black hole it was argued that a constant mode can be transmitted to both black hole and cosmological horizons. 
However the constant solution carries no stress-energy. A theoretical justification for the behavior of $\ell=0$ mode is given in \cite{brady1} 
in SdS spacetime. 

For large values of $\epsilon$, the $\ell>s$ modes of scalar, electromagnetic and gravitational perturbations, still show a power-law 
decay, having a slower decay rate than the corresponding Schwarzschild case. As the value of $\epsilon$ decreases, the power-law decay 
gives way to an exponential decay. The results for the SdS black hole spacetime, the extreme case of quintessence are consistent with the previous 
studies\cite{brady1,brady2,molina}. 

Our results may give insights to the issues related to the instability of Cauchy horizon of the black hole. 
For instance, it was demonstrated in \cite{brady4} that for all values of the physical parameters, the 
Cauchy horizon inside charged black holes embedded in de Sitter spacetime is unstable to linear perturbations. 
In view of the above results, it deserves a further detailed 
study on the radiative tails of perturbations around the charged black hole in the quintessence model which is under investigation.

\section*{Acknowledgment}
NV wishes to thank UGC, New Delhi for financial support under RFSMS scheme. VCK is thankful to CSIR, New Delhi for financial support under 
Emeritus Scientistship scheme and wishes to acknowledge Associateship of IUCAA, Pune, India.

\end{document}